\documentclass[conference]{IEEEtran}
\IEEEoverridecommandlockouts
\usepackage{cite}
\usepackage{amsmath,amssymb,amsfonts}
\usepackage{graphicx}
\usepackage{textcomp}
\usepackage{xcolor}
\usepackage{braket}
\usepackage{caption}
\usepackage{subcaption}
\usepackage{svg}
\usepackage{algorithm2e}
\RestyleAlgo{ruled}
\usepackage{algpseudocode}
\def\BibTeX{{\rm B\kern-.05em{\sc i\kern-.025em b}\kern-.08em
    T\kern-.1667em\lower.7ex\hbox{E}\kern-.125emX}}
\begin{document}

\long\def\authornote#1#2{%
  \leavevmode\unskip\raisebox{-3.5pt}{\rlap{\textcolor{#2}{$\scriptstyle\diamond$}}}%
  \marginpar{\raggedright\hbadness=10000
    \def\baselinestretch{0.8}\tiny
    \it \textcolor{#2}{#1}\par}}

\newif\ifcomments
\commentstrue 
\newcommand{\tor}[1]{\ifcomments\authornote{{\bf Tor:} #1}{red}\fi}
\newcommand{\olivia}[1]{\ifcomments\authornote{{\bf Olivia:} #1}{blue}\fi}
\newcommand{\gideon}[1]{\ifcomments\authornote{{\bf Gideon:} #1}{green}\fi}

\title{Rotation-inspired circuit cut optimization\\
{}
}

\author{\IEEEauthorblockN{Gideon Uchehara, Tor M. Aamodt, Olivia Di Matteo}
\IEEEauthorblockA{\textit{Electrical and Computer Engineering} \\
\textit{University of British Columbia}\\
Vancouver, Canada \\
\{ gideon.uchehara, aamodt, olivia \}@ece.ubc.ca}
}

\maketitle

\begin{abstract}
Recent works have demonstrated that large quantum circuits can be cut and decomposed into smaller clusters of quantum circuits with fewer qubits that can be executed independently on a small quantum computer. Classical post-processing then combines the results from each cluster to reconstruct the output of the original quantum circuit. However, the runtime for such hybrid quantum-classical algorithms is exponential in the number of cuts on a circuit. We propose Rotation-Inspired Circuit Cut Optimization (RICCO), an alternative method which reduces the post-processing overhead of circuit cutting, at the cost of having to solve an optimization problem. RICCO introduces unitary rotations at cut locations to rotate the quantum state such that expectation values with respect to one set of observables are maximized and others are set to zero. We demonstrate practical application of RICCO to VQE by classically simulating a small instance of VQE and comparing it to one of the existing circuit-cutting methods.
\end{abstract}

\section{Introduction}\label{Introduction}

 Simulating quantum circuits with a large number of qubits is intractable on a classical computer. Also, it is difficult to program these circuits on actual quantum hardware due to size constraints (insufficient qubits) and because they are also error-prone. To increase the use of current noisy small-scale quantum devices, methods have been developed to combine classical and quantum computers to simulate quantum circuits with a large number of qubits \cite{peng, tang, tomography, xanadu}.

For example, Peng et al.~\cite{peng} demonstrated that a quantum circuit can be divided into subcircuits, each with fewer qubits, which can then be run separately on a quantum computer too small to run the original circuit. The quantum computer's results are sent to a classical computer, which combines them to replicate the expected output of the original quantum circuit. The number of quantum measurements and traditional post-processing involved in replicating the original quantum circuit increases with the number of cuts. Specifically, the cost of executing the original quantum circuit is exponential in the number of cuts. Tang et al.~\cite{tang} proposed CutQC, 
a framework that employs heuristics to select a cut  reducing classical post-processing overhead. To reduce classical post-processing resources necessary to characterize circuit cutting, Perlin et al.~\cite{tomography} proposed maximum-likelihood fragment tomography (MLFT). More recently, Lowe et al.~\cite{xanadu} employed a randomized measurements method to speedup quantum circuit cutting. We refer to these procedures generically as \emph{quantum circuit cutting}.


In this paper we propose Rotation-Inspired Circuit Cut Optimization (RICCO), a method to reduce the cost of simulating a large quantum circuit on a small quantum computer. RICCO introduces a parameterized unitary operator and its adjoint at each cut location. After optimization, the unitary operator rotates a given quantum state such that it maximizes the expectation value of select Pauli observables. This is similar to rotating a quantum state vector such that it aligns with one of the Bloch sphere axes for a 1-qubit system. By this method, we effectively set the expectation value of the select Pauli observable(s) to maximum while the others are set to zero. This is relevant because we have eliminated the need to measure in multiple bases, as is the case in the original method, resulting in fewer circuit executions. We demonstrate through simulation that when compared to Peng et al.~\cite{peng} RICCO reduces the total number of quantum circuit measurements required to reconstruct a cut circuit, at the expense of the optimization.

\begin{figure}[htbp]
\centerline{\includegraphics[scale=0.25]{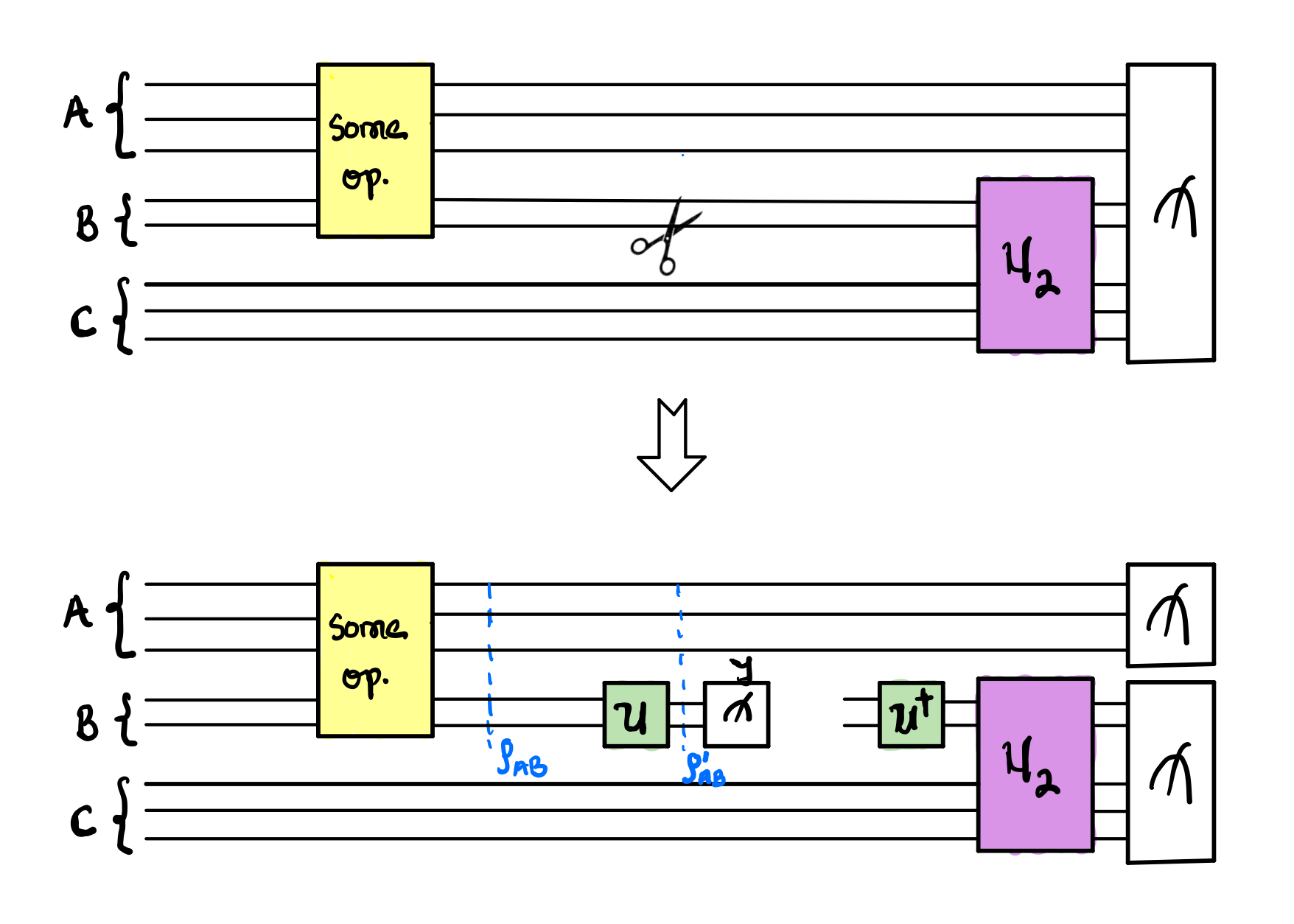}}
\caption{Quantum circuit with unitary operator $U$ applied at the cut location to rotate the state $\rho_{AB}$ to $\rho^{\prime}_{AB}$}
\label{fig: circuit}
\end{figure}

\begin{figure*}[htbp]
     \centering
     \begin{subfigure}[b]{0.3\textwidth}
         \centering
         \includegraphics[width=1.2\textwidth]{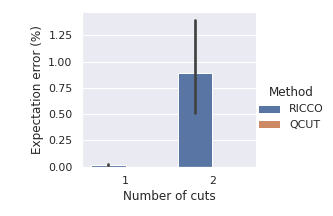}
         \caption{Expectation value error of the observable $Z^{\otimes n} = Z^{\otimes (n_A + n_B + n_C)}$ for random circuits. The error bar indicates a deviation of less than 0.3\% for two qubit cuts}
         \label{fig:ricco_cutqc_expectation_error}
     \end{subfigure}
     \hfill
     \begin{subfigure}[b]{0.3\textwidth}
         \centering
         \includegraphics[width=0.92\textwidth]{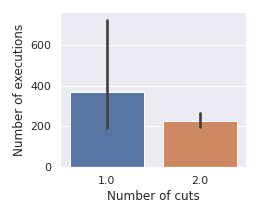}
         \caption{Circuit executions during optimization. The error bar is high because of the variance in the class of quantum circuits optimized using RICCO}
         \label{fig:ricco_circuit_opt_execution}
     \end{subfigure}
     \hfill
     \begin{subfigure}[b]{0.28\textwidth}
     \centering
     \includegraphics[width=1.2\textwidth]{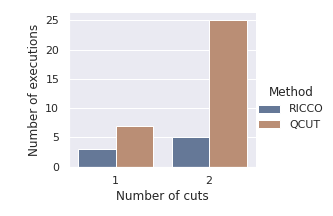}
\caption{The number of executions of quantum circuits for RICCO is significantly less than that for QCUT after optimization.}
\label{fig:ricco_cutqc_executions}
     \end{subfigure}
        \caption{Empirical results showing the performance of RICCO vs. QCUT for quantum circuits formed from random unitaries as the circuit in 
        Fig. \ref{fig: unitaries}.}
        \label{fig:errors}
\end{figure*}

\section{Theory}\label{Theory}
Consider the quantum circuit in Fig. \ref{fig: circuit}. A cut is applied on the qubits labeled $B$ resulting in two separate subcircuits: upstream and downstream subcircuits. The ultimate goal is to combine the two subcircuits to reconstruct the output of the original uncut circuit. Our focus at the moment will be on the upstream subcircuit with the qubit register labeled $A$ and the parts of qubit register $B$ just before the cut. The downstream subcircuit consists of the part of qubit register $B$ just after the cut and beyond and the qubit register labeled $C$.

In prior circuit cutting methods, measurements are made in multiple bases at the cut point, and specific eigenstates prepared to subsequently run the downstream circuits. This results in an exponentially large number of measurements and circuits. The goal of our proposed method is to find a way to reduce the number of measurements.

Without loss of generality, we are interested in computing the expectation value of some observable, say, $Z^{\otimes n} = Z^{\otimes (n_A + n_B + n_C)}$ (this also applies to observables with some Pauli $X$ and $Y$, in which case we just apply the corresponding local basis rotations) for the circuit in Fig. \ref{fig: circuit}, where $n_A$, $n_B$ and $n_C$ are the number of qubits for register labels $A$, $B$ and $C$ respectively.  To do this, we seek a unitary operator $U$ that rotates the state $\rho_{AB}$ to the state $\rho^{\prime}_{AB}$ such that $\rho^{\prime}_{AB}$ aligns with one of the computational basis vectors of the observables in $\{ \{Z^{\otimes n_A}\} \otimes \{I,Z\}^{n_B}\}$. As a result, we only have to measure in the computational basis. To find $U$, we decompose $U$ into parameterized rotations and optimize for its parameters in the quantum circuit.

To understand the intuition behind our method, consider that the density matrix $\rho$ of an arbitrary 1-qubit quantum state can be expressed in terms of Pauli matrices, 

\begin{equation}
    \rho = \frac{1}{2} \sum_i Tr(P_i \rho) P_i\label{eq1}.
\end{equation}

\noindent Here $P_i \in \{ I, X, Y, Z\}$, and $\hbox{Tr}(P_i \rho)$ is the expectation value, $\langle P_i \rangle$ of the Pauli $P_i$. By applying the unitary rotation
$U$ to $\rho$ we transform  $\rho$ to the state $\rho^\prime$ as follows:

\begin{equation}
    \rho \mapsto \rho^{\prime} = U \rho U^\dagger\label{eq2}.
\end{equation}

Since $\rho^\prime$ is also a density matrix, we can express it as 

\begin{equation}
    \rho^{\prime} = U \rho U^\dagger= \frac{1}{2} \sum_i Tr(P_i U \rho U^\dagger) P_i \label{eq3}.
\end{equation}

The first intuition behind our method is the Bloch sphere. The expectation values $\langle X \rangle$, $\langle Y \rangle$, and $\langle Z \rangle$ of the three Pauli matrices, allow one to identify the three coordinates with respect to $X$, $Y$, and $Z$ axes. If we can find a unitary operator that rotates the Bloch vector of the density matrix to align with a particular \emph{state} (e.g., the state with positive eigenvalue) of the $Z$-axis, we are by implication zeroing out the expectation values of Pauli $X$ and $Y$ observables. To avoid restricting ourselves to one eigenstate of Pauli $Z$, we can find a unitary, $U$ such that when applied to $\rho$, its final state $\rho^\prime$ ends up along either the $\ket{0}$ axis $(\langle Z \rangle = +1)$ or $\ket{1}$ axis $(\langle Z \rangle = -1)$.

The second intuition behind our method stems from the theory of optimal measurements and mutually unbiased bases (MUBs). 
MUBs are defined such that for any two vectors $\ket{\mu}$ and $\ket{\nu}$ from bases $M$ and $N$,

\begin{equation}
    \vert\langle \nu \vert \mu \rangle \vert^2 = \frac{1}{d},
\end{equation}

\noindent where $d$ is the dimension of the system. MUBs for an $n$-qubit system have a standard construction based on the partitioning of the Pauli group into $2^n + 1$ disjoint sets of $2^n - 1$ commuting operators whose mutual eigenvectors form the measurement bases \cite{mubs}. If we align our state with an eigenvector of one such basis, it will have equal overlap with all vectors in the other sets. This has further implications for the expectation values of Paulis: the state will have a non-zero expectation value for all Paulis in that set we measure with respect to, while for Paulis from other sets, they will be 0. This enables us to perform only a single measurement in the $Z$ basis.

\subsection{Cost Function to Determine the Unitary Operator}
Let's consider the case where $\rho$ is a single-qubit state. We want to rotate the state such that we point along either the $\ket{0}$ or $\ket{1}$ axis of the Bloch sphere; these are eigenstates of only Pauli $Z$, so we can ignore any observable, $P_i \notin \{I, Z\}$. The identity operator is kept because $\ket{0}$ and $\ket{1}$ are also the eigenstates of $I$.

To design the right cost function, that optimizes the parameters of a unitary operator $U$ such that it rotates $\rho$ to a new state $\rho^{\prime}$, we can use a measure of distance such as fidelity ~\cite{nielsen}:

\begin{equation}
     F \left (\ket{\psi}, \rho^{\prime} \right ) = \bra{\psi} \rho^{\prime} \ket{\psi} \label{eq4},
\end{equation}

\noindent to determine how close $\rho^{\prime}$ is to our desired state $\ket{\psi}$. We decided to use fidelity in this case because it has a straightforward definition for the case of pure versus mixed state. We want to see how close our transformed mixed state $\rho^\prime$ from the cut is to the desired pure state (a computational basis state). The best possible outcome is $F=1$, when $\rho^{\prime}$ is the same state as the desired state $ \ket{\psi} \bra{\psi}$. The worst is $F=0$ when the desired state is perfectly orthogonal to $\rho^{\prime}$.

\begin{figure*}[htbp]
     \centering
     \begin{subfigure}[b]{0.49\textwidth}
         \centering
         \includegraphics[width=\textwidth]{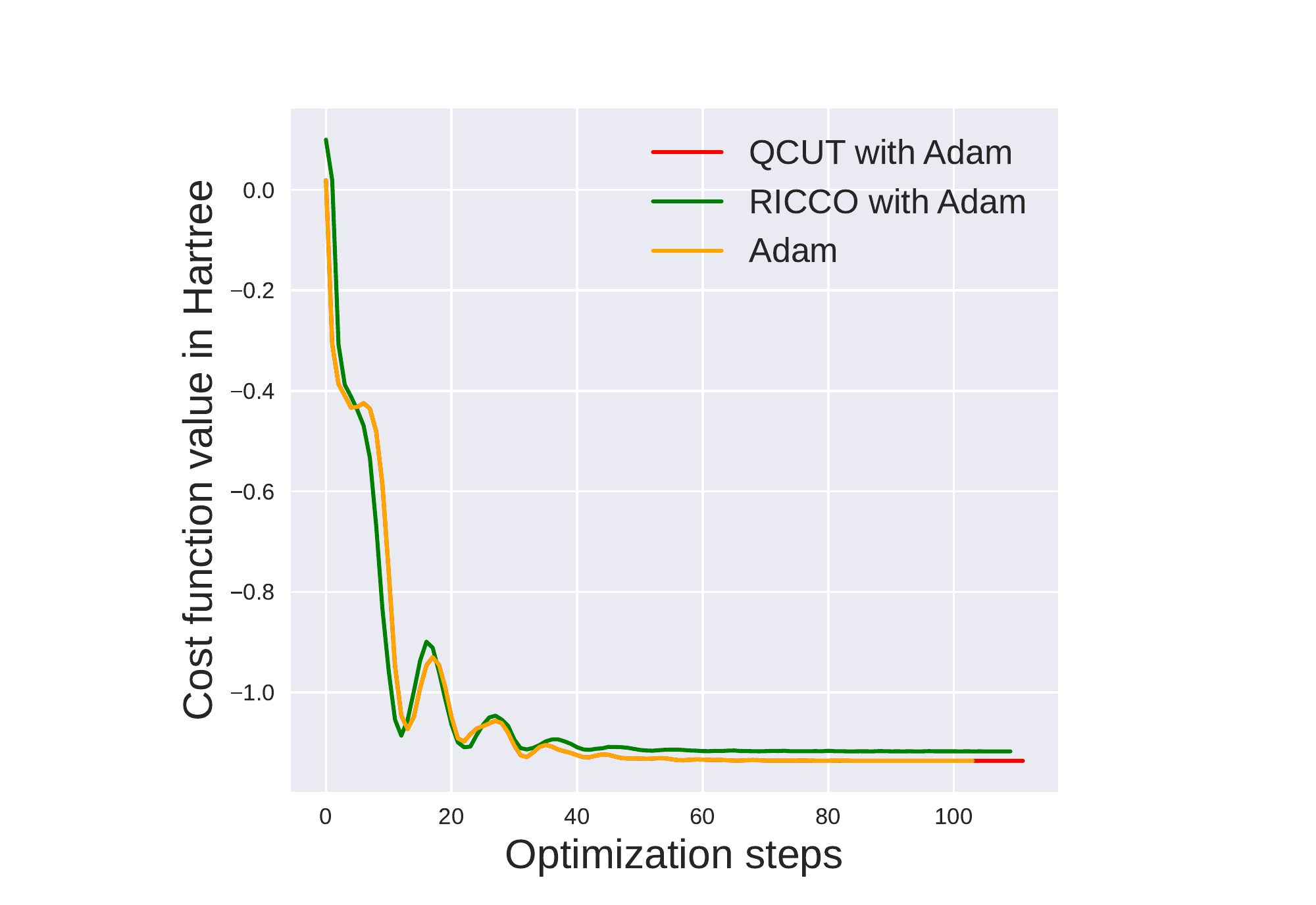}
         \caption{VQE cost function}
         \label{fig:vqe_Cost_function}
     \end{subfigure}
     \begin{subfigure}[b]{0.49\textwidth}
         \centering
         \includegraphics[width=0.95\textwidth]{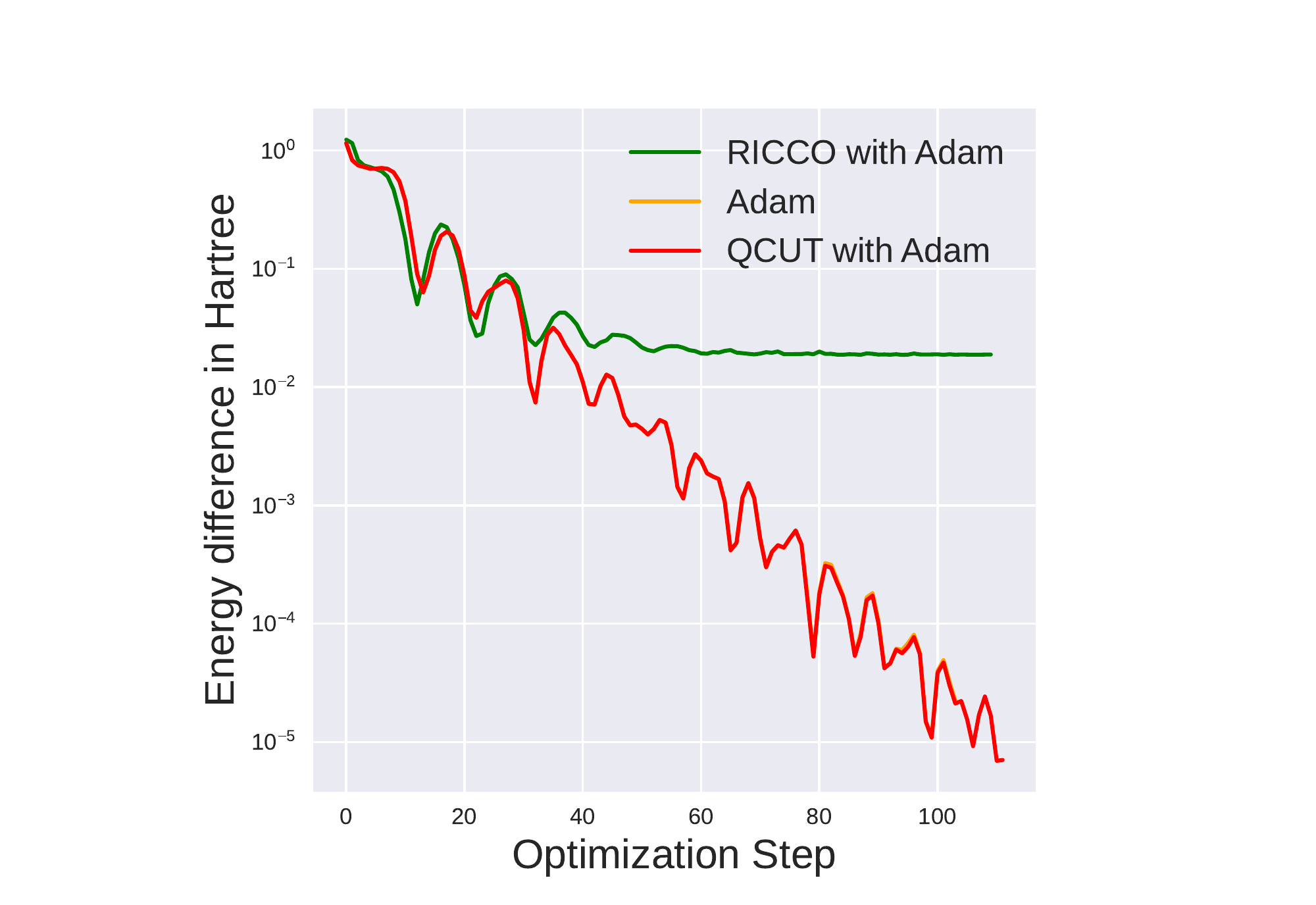}
         \caption{Energy difference curve}
         \label{fig:energy_difference}
     \end{subfigure}
        \caption{Training curve for VQE with RICCO, QCUT and conventional VQE methods. ~\ref{fig:vqe_Cost_function} shows that RICCO's performance was similar to other methods. ~\ref{fig:energy_difference} shows the deviation of the ground state energy for the different methods compared with the actual ground state energy. RICCO converges to near the true energy, but with a larger energy difference with the true value than other methods.}
        \label{fig:vqe_training}
\end{figure*}

Another nice property of fidelity is that it can be expressed in terms of expectation values as~\cite{fidelity_main, fidelity} 

\begin{equation}
\begin{split}
     F\left (\ket{\psi}, \rho^{\prime} \right ) &= \bra{\psi} \left ( \frac{1}{2} \sum_i Tr(P_i U \rho U^\dagger) P_i \right) \ket{\psi}, \\
     &=  \frac{1}{2} \sum_i Tr(P_i \rho^{\prime}) \bra{\psi} P_i \ket{\psi},\\
     &=  \frac{1}{2} \sum_i \langle P_i \rangle_{\rho^{\prime}} \langle P_i \rangle_{\ket{\psi}}.
     \label{eq5}
\end{split}
\end{equation}

\noindent These expectation values are what we measure in RICCO (see Algorithm~\ref{alg:one}). To determine $U$, we have to build a cost function that maximizes the fidelity with respect to either the $\ket{0}$ or $\ket{1}$ state. We propose to use the cost function

\begin{equation}
    C \left (\rho^{\prime} \right ) =  [1 - F \left (\ket{0}, \rho^{\prime} \right )   F \left (\ket{1}, \rho^{\prime} \right )]^2\label{eq6}
\end{equation}

\noindent This cost function is simply the squared difference between 1 and the product of the fidelities of $ \ket{0}$ and $\ket{1}$ with $\rho^{\prime}$. We squared the cost function to ensure that it is a convex function for optimization purposes.

Thus, for the one qubit case, our cost function is minimized when $\langle Z \rangle_{\rho^{\prime}} = \pm 1$. It is maximized when $\langle Z \rangle_{\rho^{\prime}} = 0$, which corresponds to states along the other two axes $X$ and $Y$. Thus, by minimizing this cost function, we are effectively setting the expectation values of $X$ and $Y$ to zero. 

For the general case, to zero out as many expectation values as possible, we need to rotate only the qubits being cut so that $\rho^\prime$ aligns with any of the computational basis vectors. Then we can measure the upper subcircuit in only the computational basis. It is important to note that as a result of optimization error, the other uncut qubits of $\rho$ may not align perfectly with the computational basis vectors of their observables. Our results show that this error is negligible in most cases and does not constitute a major limitation in reconstructing the original expectation value. For $n$ qubits, we can express the density matrix of an arbitrary state $\rho^{\prime}$ as in \eqref{eq11},

\begin{equation}
    \rho^{\prime} = \frac{1}{2^n} \sum_{P_i \in \{I, X, Y, Z\}^n} Tr(P_i \rho^{\prime}) P_i.\label{eq11}
\end{equation}

The formula for the fidelity is the same, except now we must consider all observables that consist only of $Z$s and $I$s for the cut qubit. The uncut qubits still retain their respective Pauli observables. For $n_B$ cut qubits, the observables are in the set $\{I, Z\}^{n_B}$. The fidelity of the transformed state $\rho^{\prime}$ and one of the computational basis states $\ket{\psi}$ of the combined observables $P_i \in P$ (where $P =  \{ \{Z^{\otimes n_A}\} \otimes \{ I, Z\}^{n_B}\}$) of the cut qubits, $n_B$ and the uncut qubits, $n_A$ is 

\begin{equation}
\begin{split}
     F\left (\ket{\psi}, \rho^{\prime} \right )
     &=  \frac{1}{2^n} \sum_{P_i \in P} \langle P_i \rangle_{\rho^{\prime}} \langle P_i \rangle_{\ket{\psi}}
     \label{eq12}
\end{split}
\end{equation}

\begin{algorithm}
\caption{RICCO algorithm}\label{alg:one}
\hspace*{\algorithmicindent} \textbf{Input}: upstream and downstream subcircuit, observables in Hamiltonian and cost function equation ~\ref{eq6}\\ 
 \hspace*{\algorithmicindent} \textbf{Output}: $\braket{Y}$, \textbf{Expectation value} 

$\epsilon \gets 10^{-7}$ \text{set the tolerance of RICCO}\;
$t \gets 1$ \text{set the initial tolerance}\;
\text{Initialize $X$} \text{(parameters of $U$)}\;
\text{Initialize $cost_{previous}$}\;
\For{\text{observable in optimization observables}}{
    \text{Initialize $params$}\;
    \While{$t > \epsilon$}{
    $params$, $cost_{new} \gets$ \text{Optimize the cost function in equation ~\ref{eq6}}\;
    
    $t \gets \lvert cost_{previous} - cost_{new} \rvert$ \;
    }
    $X \gets params$\;
}
$E \gets$ measurement of expectation value of subcircuits with optimized $X$\;
$\braket{Y} \gets $ combining results of subcircuits' expectation values using tensor product of corresponding observables\; 
\Return $\braket{Y}$
\end{algorithm}

Finally, we use these to construct our cost function. Let $\ket{\Bar{x}}$, for $\Bar{x} \in \{ 0, 1 \}^n$ denote our $n$-qubit computational basis state. The function to minimize is 
\begin{equation}
\begin{split}
    C_n(\rho^{\prime}) &= \left ( 1 - \prod_{\Bar{x} \in \{ 0, 1 \}^n} F\left (\ket{\Bar{x}}, \rho^{\prime} \right ) \right)^2\\
     &= \left ( 1 - \left ( \frac{1}{2^n} \right )^{2^n} \prod_{\Bar{x} \in \{ 0, 1 \}^n} \sum_{P_i \in P} \langle P_i \rangle_{\rho^{\prime}} \langle P_i \rangle_{\ket{\Bar{x}}} \right)^2
     \label{eq13}
\end{split}
\end{equation}

Whenever we rotate the state $\rho$ and align it with one of the computational basis states, the fidelities with respect to all others will be minimized. Furthermore, based on our choice of basis state, we can reconstruct the original expectation value for the observable of interest with only one measurement setting in the computational basis.

\section{Experimental Results} \label{Experiments}
We evaluated RICCO by comparing the total number of quantum circuit executions after optimization against the method by Peng et al. ~\cite{peng}, which we denote as QCUT. Because of the available computing resources, we simulated a total of 220 quantum circuits. The circuits were composed of two randomly-selected operations from unitary groups $U(16)$ and $U(32)$  for 1-qubit cut and 2-qubit cuts respectively. This is to ensure that we tested our method on the worst case possible. The total number of qubits for each circuit was 7 and  8 for 1-qubit cut and 2-qubit cuts respectively.  Simulations and optimization (using Adam ~\cite{adam}) were performed using PennyLane~\cite{pennylane}, an open-source Python software framework for quantum differentiable programming.

\begin{figure}[htbp]
\centerline{\includegraphics[scale=0.35]{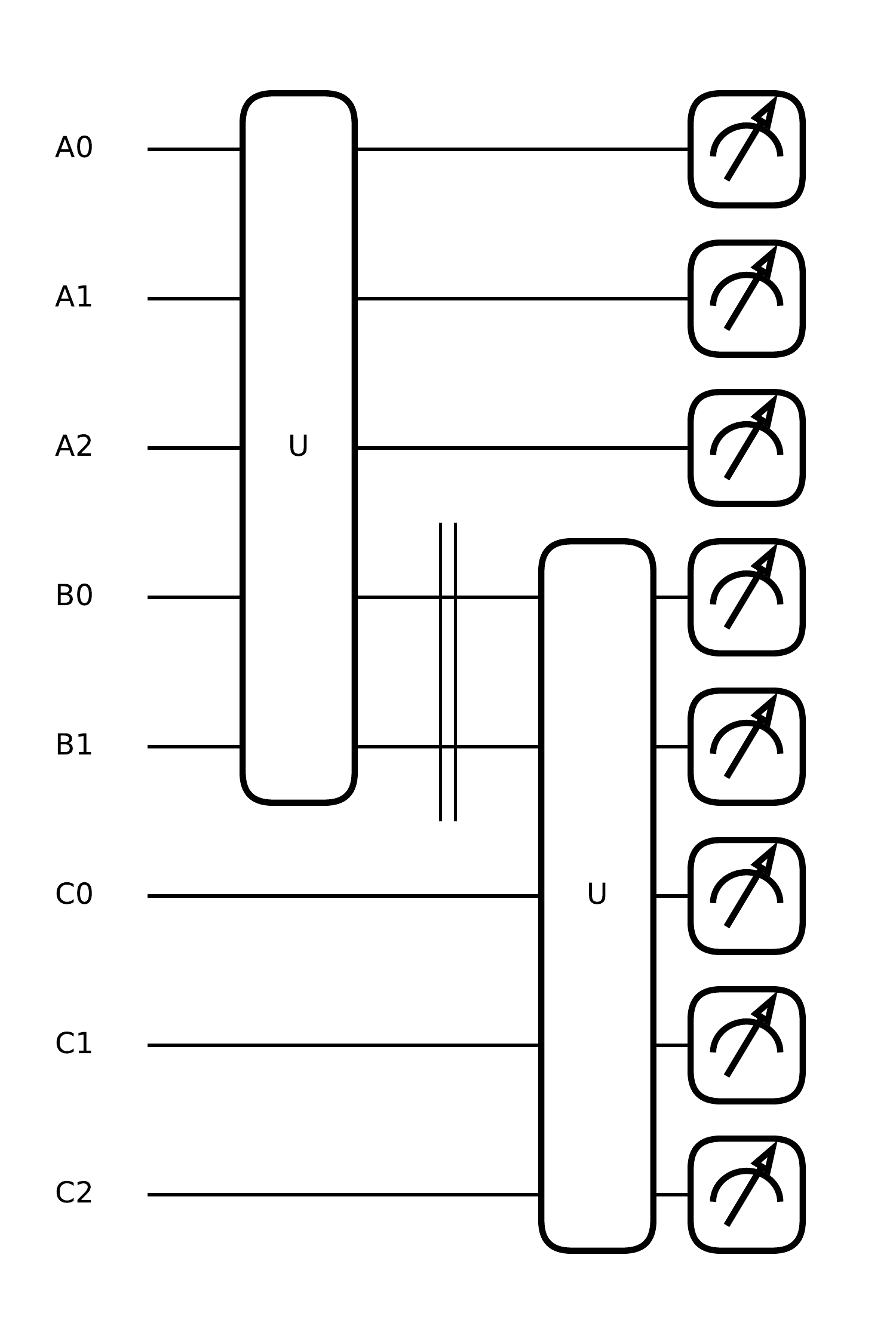}}
\caption{Example of one of the quantum circuits formed by the combination of two unitaries from the $U(32)$ unitary group. The double line indicates the cut location on two qubits}
\label{fig: unitaries}
\end{figure}

Fig. ~\ref{fig:ricco_cutqc_expectation_error} plots the percentage error of the expectation values $Z^{\otimes n} = Z^{\otimes (n_A + n_B + n_C)}$ of RICCO and QCUT circuit cutting methods compared with the ideal classical simulation of the uncut circuit for different number of cut qubits. This plot gives an insight into the accuracy of our method. We observed a reasonably high accuracy in RICCO with the worst case having less than $1.5\%$ error. Fig.~\ref{fig:ricco_circuit_opt_execution} plots the number of quantum circuit executions during optimization for the parameters of $U$ in RICCO for 1-qubit and 2-qubits cut. On average, we require about 400 quantum circuit executions during optimization for the 15 parameters of $U$ in the case of a 2-qubit cut. This number is very high partly because of the type of optimizer used. A better optimizer could reduce the number of executions required. We believe that more work needs to be done to explore how the optimization can be improved.

Fig. \ref{fig:ricco_cutqc_executions} plots the number of executions required to reconstruct the output of the uncut quantum circuit for RICCO and QCUT versus the number of cuts after optimization of the parameters of $U$. For circuits with 1 cut, the total number of quantum circuit executions for RICCO was 3 while QCUT was 7; and for 2 cuts, the total number of quantum circuit executions for RICCO was 5 while QCUT was 25. This result shows that, after optimization, RICCO improves run-time compared to QCUT.

\begin{figure*}[htbp]
\centerline{\includegraphics[scale=0.450]{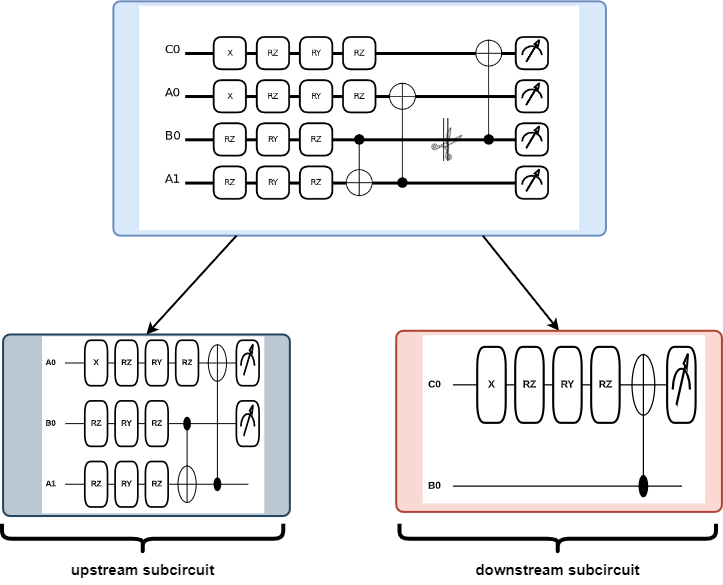}}
\caption{VQE circuit ansatz for the simulation of hydrogen molecule. The double line with scissors on register $B0$ indicates the cut location. Two subcircuits emerge after the cut: the upstream and downstream subcircuits. The upstream subcircuit consists of qubit $A0$, the part of qubit $B0$ just before the cut and $A1$. The downstream subcircuit consists of qubit $C0$ and the part of qubit $B0$ just after the cut location}
\label{fig:vqe_circuit}
\end{figure*}

\begin{figure*}[htbp]
     \centering
     \begin{subfigure}[b]{0.45\textwidth}
         \centering
         \includegraphics[width=0.8\textwidth]{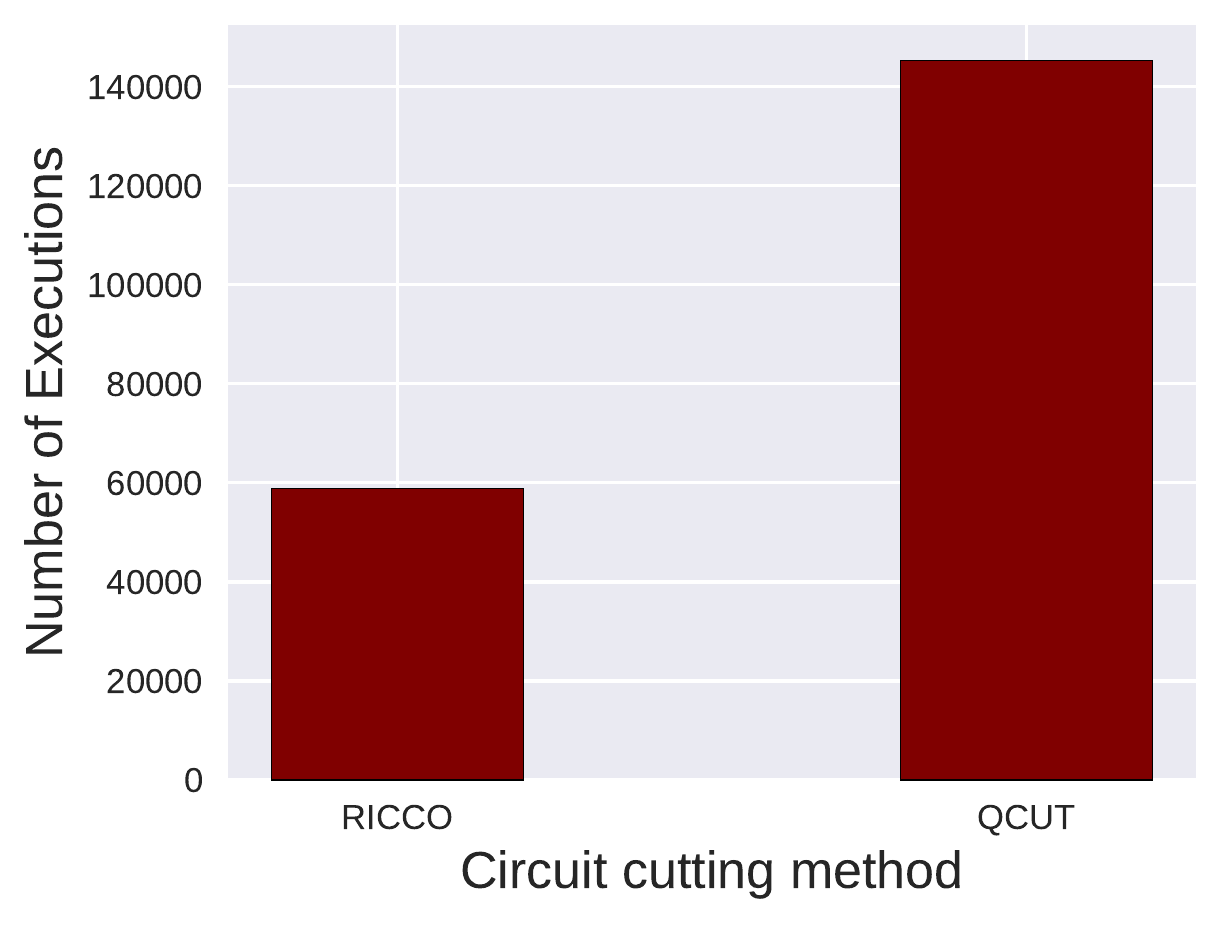}
         \caption{Number of executions for RICCO and QCUT with VQE}
         \label{fig:number_of_execution}
     \end{subfigure}
     \hfill
     \begin{subfigure}[b]{0.45\textwidth}
     \centering
     \includegraphics[scale=0.5]{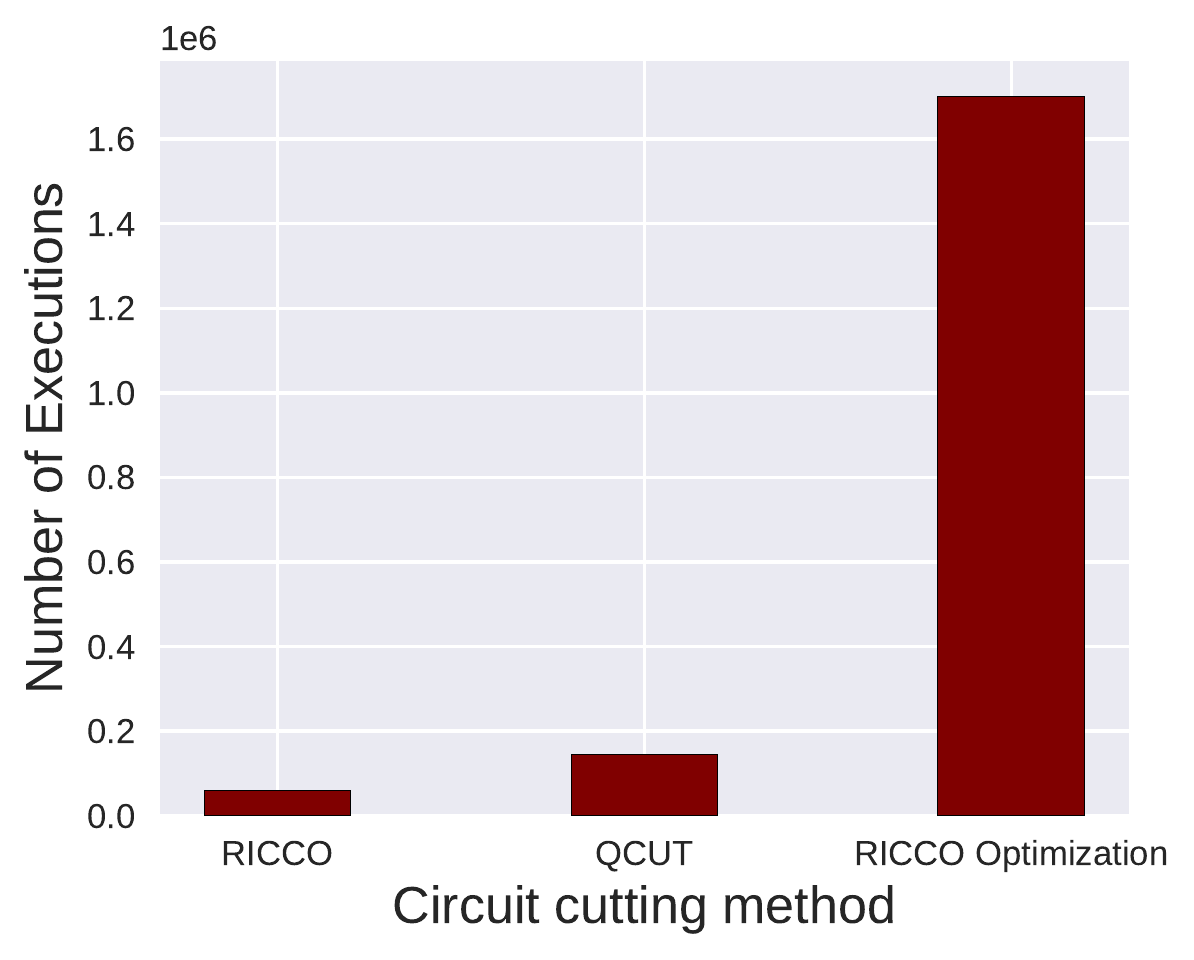}
\caption{Number of executions for RICCO Optimization, RICCO and QCUT with VQE}
\label{fig:total_number_of_executions}
     \end{subfigure}
        \caption{Optimization Results for VQE. ~\ref{fig:number_of_execution} shows that RICCO requires less than half the number of quantum circuit executions compared to QCUT. ~\ref{fig:total_number_of_executions} shows that RICCO's speedup was at the expense of its optimization }
        \label{fig:ricco_vqe}
\end{figure*}

\section{Circuit Cutting with VQE} \label{vqe}
After verifying correctness of the procedure and its behaviour in the worst case, it is of interest to evaluate its potential in practical applications. Variational quantum eigensolver (VQE) is one of the quantum algorithms in computational quantum chemistry used to predict the electronic structure and properties of molecules. In this section, we discuss the results of applying RICCO to perform numerical simulations of VQE to estimate the ground state energy of the hydrogen molecule.

PennyLane was used to build a representation of the electronic Hamiltonian. 
We used the circuit from the PennyLane demo by Li et al. ~\cite{demo} as our ansatz (see Fig. ~\ref{fig:vqe_circuit}). The parametrized circuit prepares the trial quantum state of $H_2$ molecule which is trained using a cost function that measures the expectation value of the problem Hamiltonian in the trial state to compute the ground state energy. The paramters of VQE were randomly initialized at the start of optimization with Adam's optimizer.

\begin{algorithm}
\caption{VQE with RICCO algorithm}\label{alg:two}
\hspace*{\algorithmicindent} \textbf{Input}: upstream and downstream subcircuit, RICCO subroutine, and $vqe_{params}$ \\ 
 \hspace*{\algorithmicindent} \textbf{Output}: $\braket{Y}$, \textbf{Expectation value} 

$\epsilon \gets 10^{-6}$ \text{set the tolerance of VQE}\;
$max_{iter} \gets 500$ \text{set the maximum number of iterations}\;
$iter \gets 0$ \text{initial iterations}\;
$t \gets 0$\;
\text{Initialize X}\; \text{parameters of U}
\text{Initialize $cost_{previous}$}\;
    
    \text{Initialize $params$}\;
    \While{$iter < max_{iter}$}{
       $ricco_{params}$ $\gets$ \text{RICCO optimization}\;
        
       VQE Optimization step $\gets$ $ricco_{params}$\;
        
        ${vqe}_{params}, {vqe}_{energy} \gets \text{VQE Optimization step}$\;
        
        $t \gets \lvert prev_{energy} - vqe_{energy} \rvert$ \;
        
        $cost \gets vqe_{energy}$\;
    
        \If{$t \leq \epsilon$}{
            {\textbf{break}}
        }
    }
\Return $cost$
\end{algorithm}

To apply RICCO to VQE, one of the qubits of the original ansatz was cut which resulted into two subcircuits before training. During the training phase of the VQE for the cut circuit, RICCO was used to reconstruct the expectation value for the cost function (see Algorithm ~\ref{alg:two}).

The Hamiltonian consists of 18 observables which contribute to its expectation value. After cutting the circuit, it is expected that each of these observables are treated as separate entities and optimized separately with RICCO (see Algorithm ~\ref{alg:one}). The results of the expectation value of each observable are then recombined to reconstruct the original expectation value of the uncut circuit. To reduce the number of RICCO optimization and hence, the number of executions for the 18 observables, we grouped them into 6 disjoint sets of observables, \{$\{III, IZI\}$, $\{IIZ, IZZ\}$, $\{XIY, XZY\}$, $\{YIX, YZX\}$, $\{ZII, ZZI\}$, $\{ZIZ, ZZZ\}$\} for optimal RICCO optimization of the upstream subcircuit. This is an optimal choice because it  captures all the observables within the Hamiltonian that belong to the uncut qubits of the upstream subcircuit and avoids repeating optimization for observables that are the same on the uncut qubits. It is important to note that for each observable in a set, only the middle observable changed -- this is the cut qubit, $B0$. The first and last observables are the qubit registers $A0$ and $A1$ respectively. 

At every step of VQE optimization, RICCO was deployed to reconstruct the original expectation value of the Hamiltonian. The set of 6 unique observables also consists of non-Z observables, $\{XIY, XZY\}$ and $\{YIX, YZX\}$ that were also optimized  separately with RICCO at each step of the VQE optimization. It is important to do this to ensure the accurate reconstruction of the original expectation value. This process is repeated until VQE converges to a preset tolerance of about $10^{-6}$. It took more than 70 iterations for RICCO to converge to this tolerance as shown in Fig. ~\ref{fig:vqe_Cost_function}.

Fig. ~\ref{fig:vqe_Cost_function} compares VQE cost function versus the number of iterations for RICCO, QCUT and the conventional VQE optimization methods. Adam optimizer was used for the different methods. Results from the optimization showed that although RICCO coverged, it did not converge to the actual value with the other two methods.  This is evident in Fig. ~\ref{fig:energy_difference} which shows that the difference in RICCO's energy compared with the true value was worse (about $10^{-2}$ Hartree) than QCUT and conventional VQE with no cutting (about $10^{-6}$ Hartree). RICCO's convergence threshold was set to $10^{-7}$. Increasing this threshold indicated some improvements in the energy difference at the expense of increased circuit executions and runtime. We noted that large difference in RICCO's energy could be as a result of the errors incurred during RICCO's optimization and the convergence threshold set for RICCO. Notwithstanding, exploring other causes of this error and how to mitigate it would be the focus of our future work. 

\section{Discussion} \label{discussion}
We want to understand how to reduce the number of measurements in cutting procedures, and see if this translates to improvements in actual applications such as VQE.  Fig. ~\ref{fig:number_of_execution} shows the number of executions required to implement VQE with RICCO and QCUT. RICCO requires less than half the number of quantum circuit executions compared to QCUT. Fig. ~\ref{fig:total_number_of_executions} on the other hand, shows that number of executions required for RICCO optimization was at least $8X$ more than the number of executions for QCUT. This shows the trade-off between the improved runtime for RICCO and number of executions required for its optimization. We observed that this increase in the number of executions was because for each iteration of VQE, RICCO was optimized. One way to solve this problem could be to combine RICCO optimization with VQE optimization for each iteration of VQE instead of running a separate optimization for each. We can also reseed RICCO parameters at each optimization step.

Finally, we observed that the order of optimization of the 6 unique sets of observables could have contributed to the increased number of executions for RICCO. This is because, the optimization of any set of the observables depends on the outcome of the optimized parameters for the previous set of observable that was optimized by RICCO. Alternative distance measures such as trace distance can be explored as an area for future work to determine how they affect optimization convergence

\end{document}